\documentclass[a4paper,traditabstract]{aa}
\usepackage{graphicx}
\usepackage{txfonts}
\usepackage{color}
\usepackage{natbib}
\bibpunct{(}{)}{;}{a}{}{,} % to follow the A&A style

\def\ao{Applied Optics}

\def\apjl{Astrophys. J. Letters}

\def\apjs{Astrophys. J. Suppl.}

\begin{document}
%\begin{multicols}{1}
%\linenumbers

\title{The extrasolar planet Gliese 581 d: a potentially habitable planet? (Corrigendum)}
\titlerunning{GL 581 d habitable planet}

\author{P. von Paris\inst{1} \and S. Gebauer\inst{2} \and M. Godolt\inst{2}\thanks{now at Institut f\"{u}r Planetenforschung, Deutsches Zentrum
f\"{u}r Luft- und Raumfahrt, Rutherfordstr. 2, 12489 Berlin, Germany} \and J.~L. Grenfell\inst{2}
\and P. Hedelt\inst{1}\thanks{now at Institut f\"{u}r Methodik der Fernerkundung, Deutsches Zentrum f\"{u}r Luft-und Raumfahrt,  Oberpfaffenhofen, 82234 We{\ss}ling, Germany} \and D. Kitzmann\inst{2} \and A.~B.~C.
Patzer\inst{2} \and H. Rauer\inst{1,2} \and B. Stracke\inst{1}}

\institute{Institut f\"{u}r Planetenforschung, Deutsches Zentrum
f\"{u}r Luft- und Raumfahrt, Rutherfordstr. 2, 12489 Berlin, Germany
\and Zentrum f\"{u}r Astronomie und Astrophysik, Technische
Universit\"{a}t Berlin, Hardenbergstr. 36, 10623 Berlin, Germany}

\abstract{We report here that the equation for H$_2$O Rayleigh scattering was incorrectly stated in the original paper. Instead of a quadratic dependence on refractivity $r$, we accidentally quoted an $r^4$ dependence. Since the correct form of the equation was implemented into the model, scientific results are not affected.}


\maketitle

%\end{multicols}{1}

%\section{Introduction}

It was recently brought to our attention \citep{kopparapu2013} that in our original paper \citep{vparis2010gliese}, we stated an incorrect equation for the calculation of the H$_2$O Rayleigh scattering coefficient $\sigma_{\rm{ray,H_2O}}$. Equation 3 in \citet{vparis2010gliese} shows a $r(\lambda)^4$ dependence of $\sigma_{\rm{ray,H_2O}}$, where $r(\lambda)$ is the wavelength-dependent refractivity of H$_2$O. Instead, as stated in \citet{Allen1973}, it should be a $r(\lambda)^2$ dependence. Therefore, the correct equation ($\sigma_{\rm{ray,H_2O}}$ in cm$^2$) reads

\begin{equation}\label{rayleighwater}
\sigma _{\rm{ray,H_2O}}(\lambda)=4.577\times 10^{-21} \cdot
\left(\frac{6+3\cdot D}{6-7\cdot D}\right)\cdot \frac{r(\lambda)^2}{\lambda^4}
\end{equation}

where $D$ is the depolarization ratio and $\lambda$ the wavelength in $\mu$m. Our work assumes $D=0.17$ from \citet{marshall1990}. The refractivity is calculated as $r(\lambda)=0.85 \cdot r_{\rm{dry air}}(\lambda)$ \citep{edlen1966}. The refractivity of dry air ($r_{\rm{dry air}}(\lambda)=n_{\rm{dry air}}(\lambda)-1$) is obtained from an approximate formula for the refractive index $n_{\rm{dry air}}(\lambda)$ given by \citet{bucholtz1995}. With this equation, we calculate 2.6$\times 10^{-27}$\, cm$^2$ for the H$_2$O Rayleigh scattering cross-section at 0.6\,$\mu$m, close to the value of 2.32$\times 10^{-27}$\,cm$^2$ from \citet{selsis2007b} or 2.5$\times 10^{-27}$\,cm$^2$ from \citet{kopparapu2013}.

The numerical factor 4.577$\times 10^{-21}$ in Eq. \ref{rayleighwater} is derived from \citet{Allen1973} in the following way: \citet{Allen1973} states that the Rayleigh cross-section is

\begin{equation}\label{allenfact}
\sigma _{\rm{ray}}= \frac{32 \pi^3}{3 N^2} \cdot \left(\frac{6+3\cdot D}{6-7\cdot D}\right) \cdot \frac{r^2}{\lambda^4}
\end{equation}

where $\sigma _{\rm{ray}}$ is in cm$^2$ and $\lambda$ is in $\mu$m. $N$ is the number of particles per unit
volume, and we took, as stated in \citet{Allen1973}, standard temperature and pressure conditions ($T$=273.1\,K, $p$=1.013\,bar). This yielded
4.577$\times 10^{-37}$ for the wavelength-independent factor $ \frac{32\cdot \pi^3}{3\cdot N^2}$ in Eq. \ref{allenfact}. Since $N$ has units of cm$^{-3}$ and the cross section is in cm$^2$, one
must then transform $\lambda$ from $\mu$m to cm, i.e. multiply by 10$^{-4}$. To the
4$^{\rm{th}}$ power, this is 10$^{-16}$, which then results in the factor 4.577$\times 10^{-21}$, as stated
in Eq. \ref{rayleighwater}.

The correct equation (Eq. \ref{rayleighwater}) was implemented in the model code, hence the calculations of the H$_2$O Rayleigh scattering were treated correctly in the model used by \citet{vparis2010gliese}. Therefore the results reported in \citet{vparis2010gliese} are not affected.

The equation for H$_2$O Rayleigh scattering reported in \citet{kopparapu2013} (their Eq. 1) is incorrect. Hence, their statement that "the coefficient in the Rayleigh scattering cross section given in Von Paris et al. (2010) should be seven orders of magnitude smaller" \citep{kopparapu2013} is also incorrect. We have contacted the authors of \citet{kopparapu2013} about this, and they subsequently changed the online arxiv.org version (arXiv:1301.6674v2) to correct their equation and the corresponding text, however we point out that the printed journal version remains unchanged.

\bibliographystyle{aa}
\bibliography{literatur_err}

\end{document}